%Paper: hep-th/9304119
%From: roche@cptvax.in2p3.fr
%Date: Mon, 26 Apr 93 09:06:44 GMT

\magnification=1200
\catcode `@=11

\hsize 17truecm
\vsize 24truecm
\def\build#1_#2^#3{\mathrel{\mathop{\kern 0pt#1}\limits_{#2}^{#3}}}

It is currently of the greatest interest to analyze the most possible
models leading to classical string vacua, in particular those with exact
solutions. This, with the hope that some of them will provide interesting
enough vacua such that they may have some possible physical interpretations.

Such an example has been discovered within some gauged $WZNW$ coset
models [1] and exhibits blackhole-type solutions in two dimensional target
space.

Recently a new class of models has been introduced by Tseytlin [2].
These are $\sigma$-models with Minkowskian signature and the key idea was
to consider such models which have symmetric target space metric with
covariantly constant null Killing vector and are therefore finite.

In this note we plan to investigate a suggestion made by Tseytlin
himself [2] and concerning the case when the so-called ''transverse''
part\footnote{$^1$}{All along this note, we shall refer to
the terminology, to the notations and results, as published in [2].
In particular, formulas (4), (5), (6) and (7) are the formulas (11), (12),
(19) and (20) of ref.[2], respectively.} of the model possesses an $n=2$
supersymmetry with an homogeneous K\"ahler target space.

Ref.[2] considers the  line element :
$$\leqalignno{&ds^2=G_{\mu\nu} dx^{\mu}dx^{\nu}=-2dudv+f(u)
\gamma_{ij}(x)dx^idx^j\cr
&\hbox{with}\ \mu,\nu=0,1...,N,N+1\ ;\ i,j=1,...N&(1)}$$
and the $\gamma_{ij}$ corresponding to a symmetric space (constant curvature).
It is then shown/that, with a specific choice of $f(u)$, the $\sigma$-model
with (1) as target space metric is ultra-violet finite. The part of (1)
proportional to $f(u)$ is referred to as the ''transverse''
part and $u$ and $v$ are light cone coordinates. Furthermore, only the
case of a curvature of the form
$$R_{ijkl}=K/(N-1)\left(\gamma_{ik}
\gamma_{jl}-\gamma_{il}\gamma_{jk}\right);K\equiv R/N,\leqno (2)$$
qualified of maximally symmetric, is considered in order to make easier
the perturbative expansions of the various quantities necessary to the
interpretation of the model.

We shall not discuss any further Tseytlin's model except for the parts
relevant to our purpose. The finiteness of the model on a flat $2d$
background needs the condition
$$\beta^G_{\mu\nu}+D(_{\mu} M_{\nu})=0\leqno(3)$$
\footnote{$\S$}{For renormalization techniques in non flat spaces, see for
instance references [4]}

\noindent with $\beta_{\mu\nu}^G$ concerning the full $\sigma$-model
with target space metric (1) beta function and $M_{\nu}$ a vector to be
determined for each separate case in order to satisfy $(3)$.

The analysis of ref.[2] leads, among others, to the following basic results:
$$\beta_{ij}^G=\beta(f)\gamma_{ij}\ ,\qquad\hbox{with}\leqno(4)$$
$$\leqalignno{&\beta(f)=a+(N-1)^{-1}a^2f^{-1}+{N+3\over 4}(N-1)^{-2}
a^3f^{-2}+0(a^4f^{-3})\cr &a\equiv\alpha'K\ .&(5)\cr}$$
$f^{-1}$ plays the role of a coupling of the symmetric space $\sigma$-model
and satisfies
$${df\over d\tau}=\beta(f)\ ,\leqno (6)$$
with $\tau$ a kind of RG ''time'' parameter defined just after (18, ref.[2]).
Of course one has
$$f(u)=a\left(\tau+(N-1)^{-1}\log\tau+0(\tau^{-1})\right)\ ;\
\tau=\tau(u)\leqno (7)$$ when $\beta(f)$ is given by $(5)$ above.

Now we come to  the transverse part, which we recall is $N$-dimensional
in the target space.

One knows from previous studies that if this transverse part is K\"ahlerian,
it enjoys an $n=2$ supersymmetry. Moreover we know from a particular example
[3] that its beta function is exactly given by its first term (1-loop).
The example we refer to requires in addition that the transverse
space not only be K\"ahlerian but also homogeneous. Appropriate K\"ahler
manifolds of this type can be found in the literature (ref.[3] and ref.
therein).

Suppose our choice of space is such a manifold, in this case $(5)$ and $(7)$
reduce to
$$\beta(f)=Const\equiv C\leqno (8)$$
$$f(u)=C.\tau(u)\ .\leqno (9)$$
$\tau$ is well defined in each case in terms of $u$, as said before and
therefore the metric $(1)$ is given exactly by
$$ds^2=-2dudv+C.\tau\gamma_{ij}(x) dx^idx_j\leqno (10)$$
We conclude that the results $(8)-(10)$ confirm, in the specific case
considered, the conjecture made by Tseytlin in the footnote 3 of ref.[2].
At this point we have not yet identified any string vacua. However, depending
on whether the vector $M_{\nu}$ in $(3)$ is an exact gradient, the
appropriate dilaton field can be exactly determined in order to satisfy Weyl
invariance of the model which then leads to exact string tree-level vacua
represented by the resulting backgrounds.

\bigskip

\noindent{\bf REFERENCES}
\medskip
\parindent=1truecm
\item{\hbox to\parindent{\enskip [1]}\hfill}G. Mandal, A. Sengupta,
S. Wadia, Mod. Phys. Lett. {\bf A6} (1991) 1685;
\item{\hbox to\parindent{\enskip }\hfill}R. Dijkgraaf, H. Verlinde,
E. Verlinde, Nucl. Phys. {\bf B371} (1991) 269;

\item{\hbox to\parindent{\enskip }\hfill}A.A. Tseytlin, Phys. Lett.
{\bf B268} (1991) 175 and John-Hopkins Univ. Preprint JHU-TIPAC 91009;

\item{\hbox to\parindent{\enskip }\hfill}E. Witten, Phys. Rev. {\bf D44}
(1991) 314;

\item{\hbox to\parindent{\enskip }\hfill}M. Muller, Nucl. Phys. {\bf B337}
(1990) 37.

\item{\hbox to\parindent{\enskip }\hfill}S. Elitzur, A. Forge,
E. Robinovici, Nucl. Phys. {\bf B359}(1991) 581;

\item{\hbox to\parindent{\enskip }\hfill}M. Rocek, K. Schoutens, A. Sevrin,
 IAS Preprint IASSNS-HEP-91/14.

\item{\hbox to\parindent{\enskip [2]}\hfill}A.A. Tseytlin, Cambridge
Preprint DAMTP-92-26 hepth@xxx/9205058.

\item{\hbox to\parindent{\enskip [3]}\hfill}A. Morozov, A. Perelomov,
M. Shifman, Nucl. Phys. {\bf B248} (1984) 279;

\item{\hbox to\parindent{\enskip }\hfill}A. Morozov, A. Perelomov, ZhETF
Pisma {\bf 40} (1984) 38.

\item{\hbox to\parindent{\enskip [4]}\hfill}A.A. Tseytlin, Nucl. Phys.
{\bf B294} (1987) 383;

\item{\hbox to\parindent{\enskip }\hfill}I. Jack, H. Osborn, Nucl. Phys.
{\bf B343} (1990) 647 and references therein.

 \end